# A New Achievable Rate for the Gaussian Parallel Relay Channel

Seyed Saeed Changiz Rezaei, Shahab Oveis Gharan, and Amir K. Khandani
Coding & Signal Transmission Laboratory
Department of Electrical & Computer Engineering
University of Waterloo
Waterloo, ON, N2L 3G1
sschangi, shahab, khandani@cst.uwaterloo.ca

**Abstract**

Schein and Gallager introduced the Gaussian parallel relay channel in 2000. They proposed the Amplify-and-Forward (AF) and the Decode-and-Forward (DF) strategies for this channel. For a long time, the best known achievable rate for this channel was based on the AF and DF with time sharing (AF-DF). Recently, a Rematch-and-Forward (RF) scheme for the scenario in which different amounts of bandwidth can be assigned to the first and second hops were proposed. In this paper, we propose a *Combined Amplify-and-Decode Forward (CADF)* scheme for the Gaussian parallel relay channel. We prove that the CADF scheme always gives a better achievable rate compared to the RF scheme, when there is a bandwidth mismatch between the first hop and the second hop. Furthermore, for the equal bandwidth case (Schein's setup), we show that the time sharing between the CADF and the DF schemes (CADF-DF) leads to a better achievable rate compared to the time sharing between the RF and the DF schemes (RF-DF) as well as the AF-DF.

## I. Introduction

### A. Motivation

The continuous growth in wireless communication has motivated information theoretists to extend Shannon's information theoretic arguments for a single user channel to the scenarios that involve communication among multiple users. In this regard, cooperative communication in

Financial supports provided by Nortel, and the corresponding matching funds by the Federal government: Natural Sciences and Engineering Research Council of Canada (NSERC) and Province of Ontario: Ontario Centres of Excellence (OCE) are gratefully acknowledged.

which a source exploits some intermediate nodes as relays, to transmit its data to an intended destination has received significant attention during recent years. Relays can emulate distributed transmit antennas to combat the multi-path fading effect and increase the physical coverage area.

Since constructing a large-scale wireless network is very expensive, it is important to understand how to efficiently utilize the available power and bandwidth resources. The parallel relay channel is the basic building block of a general network. Here, our goal is to study and analyze the performance limits of this channel.

### *B. History*

The Relay channel is a three terminal network which was introduced for the first time by Van der Meulen in 1971 [1]. The most important capacity result of the relay channel was reported by Cover and El Gamal [2]. They proposed the Decode-and-Forward (DF) scheme based on block Markov encoding in which the relays decode the transmitted message. These authors also proposed the Compress-and-Forward (CF) strategy in which relays do not decode the message, but send the compressed received values to the destination. Zahedi and El Gamal considered two different cases of the frequency division Gaussian relay channel. They derived lower and upper bounds on the capacity of this channel, which in turn translates to upper and lower bounds on the minimum required energy per bit for the reliable transmission [3]. The authors also derived a single letter characterization of the capacity of the frequency division Additive White Gaussian Noise (AWGN) relay channel with simple linear relaying scheme [4] [5]. Recently, Cover and Young-Han Kim in [6] studied a class of deterministic relay channel and derived its capacity with the hash-and-forward and CF schemes. Marko Aleksic, Peyman Razaghi, and Wei Yu in [7] derived the capacity of a class of modulo-sum relay channels using the CF scheme of [2]. They showed that the capacity of this channel is strictly below the cut-set bound.

There are also several works on the multi-relay channel in the literature (See [8]–[16], [18]–[20], [23]). Xie and Kumar generalized the block Markov encoding scheme of [2] for a network of multiple relays [10]. Furthermore, Gastpar, Kramer, and Gupta extended the CF scheme in [2] to a multiple relay channel by introducing the concept of antenna polling in [12] and [13]. They showed that when the relays are close to the destination, this strategy achieves the antenna-clustering capacity. On the other hand, when relays are close to the source, the DF strategy can achieve the capacity in a wireless relay network [14]. In [15], Amichai, Shamai,

Steinberg and Kramer considered the problem of a nomadic terminal sending information to a remote destination via agents with lossless connections. They investigated the case that these agents do not have any decoding capability, so they must compress what is received. This case is also fully characterized for the Gaussian channel. In [16], we completely characterized the asymptotic capacity of the half-duplex Gaussian parallel relay channel with two relays using the Dirty Paper Coding scheme. Moreover, assuming successive relaying protocol, we derived the optimum input distribution for the source and relays. Recently, Salman Avestimehr, Suhas Diggavi and David Tse in [18]–[20] further studied the capacity of wireless relay networks. The authors in [18] [19], proposed a deterministic model for a multiuser communication channel and generalized the max-flow min-cut theorem from the wire-line to the wireless networks. In [20], they proposed an achievable rate for the Gaussian relay networks and showed that their achievable rate is within a constant bit (determined by the graph topology of the network) from the cut-set bound.

### *C. Contributions and Relation to Previous Works*

In this paper, we consider the Gaussian parallel relay channel with a source, a destination, and a set of relays. There is no direct link from the source to the destination. This parallel relay channel is a special case of a multiple relay network in which the source broadcasts its data to all the relays, and the relays transmit their data coherently to the destination.

Schein and Gallager introduced the parallel relay channel in [8] [9]. They considered the parallel relay channel with two relays and studied possible coding schemes for this channel. For the Gaussian case, they proposed the Amplify-and-Forward (AF) and Decode-and-Forward (DF) schemes and also another scheme based on the time sharing of those schemes. Gastpar in [11] showed that in a Gaussian parallel relay channel with infinite number of relays, the optimum coding scheme is the AF.

For many years, Schein and Gallager's achievable rate based on the time sharing between the AF and DF schemes (AF-DF) was the best known achievable scheme for the Gaussian parallel relay channel with two relays. Since then there was no reported improvement in the literature. However, more recently, Yuval Kochman, Anatoly Khina, Uri Erez, Ram Zamir in [23], proposed the Rematch-and-Forward (RF) scheme for this channel. This scheme is based on the use of analog modulo-lattice modulation (See [22]), and is used for the scenarios in which there is a bandwidth mismatch between the source-relays and relays-destination channels. Furthermore, the authors showed that the time sharing between the RF and DF scheme (RF-DF), in certain scenarios, achieves a better rate than the Schein and Gallager's scheme.

In this paper, we propose a Combined Amplify-and-Decode (CADF) scheme, when there is a bandwidth mismatch between the source-relays (Broadcast: BC) and relays-destination (Multiple Access: MAC) channels. We prove that this scheme always achieves a better rate than the RF scheme. Furthermore, we show that time sharing between the CADF and DF schemes (CADF-DF) always outperforms the RF-DF and the AF-DF.

This paper is organized as follows: The system model is introduced in section II. In section III, the CADF scheme for the bandwidth mismatch scenarios is explained. Also its achievable rate is compared with that of the traditional coding schemes as well as the RF scheme. Simulation results are presented in section IV, and section V concludes the paper.

### D. Notation

Throughout the paper, lowercase bold letters and regular letters represent vectors and scalars, respectively. And $C(x) \triangleq \frac{1}{2}\log_2(1+x)$. Furthermore, for the sake of brevity, $A_\epsilon^{(n)}$ denotes the set of weakly jointly typical sequences for any intended set of random variables.

## II. The System Model

The setup of the system model considered in this paper is similar to [23]. Here, we consider a Gaussian network which consists of a source, $M$ relays, and a destination with no direct link between the source and the destination.

Nodes $1, \cdots, M$ represent relay 1 , $\cdots$, relay $M$, respectively. The transmitted vectors from the source and the relays, and the received vectors at the relays and the destination are denoted by $\mathbf{x}_{BC}$, $\mathbf{x}_m (m = 1, \cdots, M)$ and $\mathbf{y}_m (m = 1, \cdots, M)$, and $\mathbf{y}_{MAC}$, respectively. Hence, we have

$$\mathbf{y}_m = \mathbf{x}_{BC} + \mathbf{z}_m, \quad m \in \{1, \cdots, M\}, \tag{1}$$

$$\mathbf{y}_{MAC} = \sum_{m=1}^{M} \mathbf{x}_m + \mathbf{z}_{MAC}. \tag{2}$$

where $\mathbf{z}_m$ and $\mathbf{z}_{MAC}$ are the AWGN terms. Throughout the paper, for the sake of simplicity, we consider the symmetric case in which all the AWGN terms have zero mean and the variance "1" per dimension.

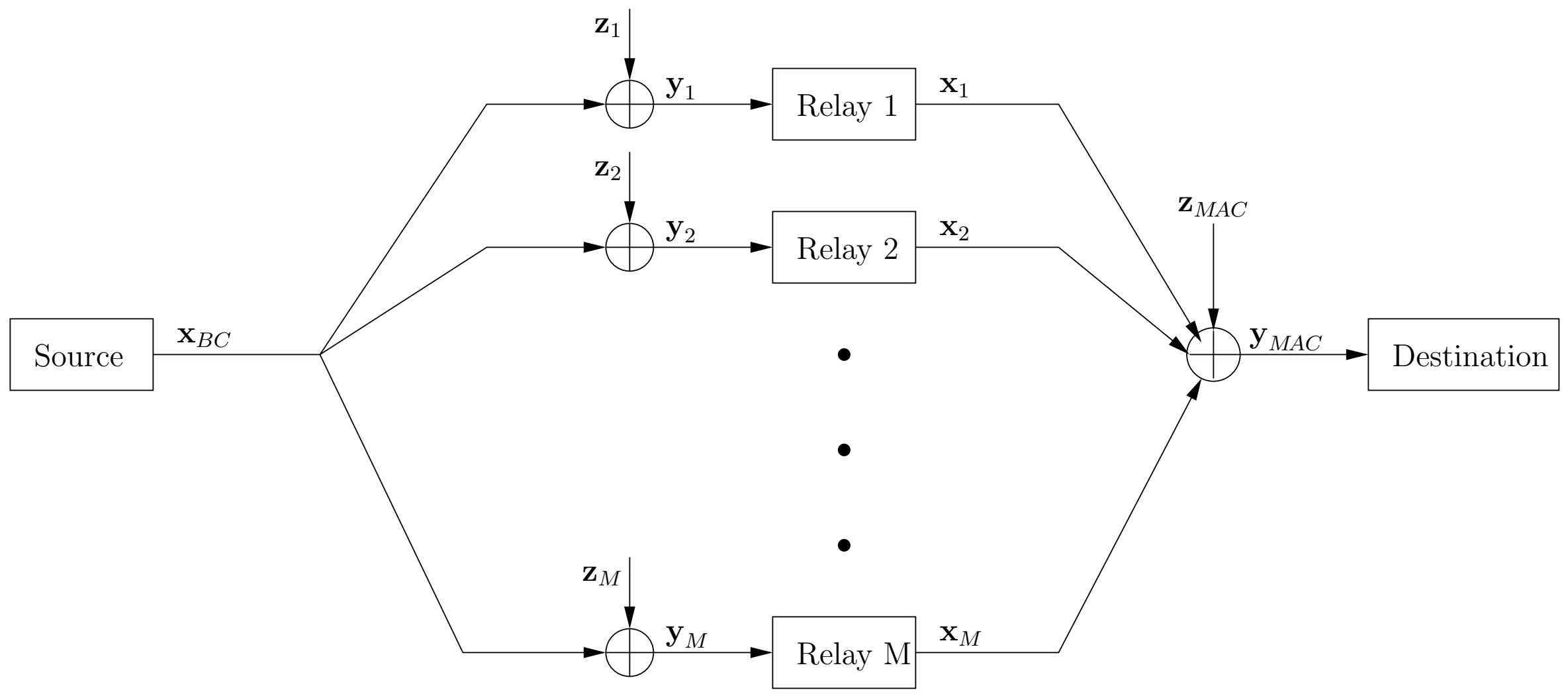

Fig. 1. The Gaussian Parallel Relay Channel.

Furthermore, the average power constraints $P_s$, $P_m$ $(m \in \{1, \cdots, M\})$ should be satisfied for the source and relay nodes:

$$\frac{1}{n} E \parallel \mathbf{x}_{BC} \parallel^2 \leq P_s, \tag{3}$$

$$\frac{1}{n} E \parallel \mathbf{x}_m \parallel^2 \leq P_m, \quad m \in \{1, \cdots, M\}. \tag{4}$$

where $n$ denotes the corresponding vector length.

Due to the symmetry assumption, we have

$$P_1 = P_2 = \cdots = P_M = P_r. \tag{5}$$

It should be noted that for the bandwidth mismatch case $P_s$ and $P_r$ are the power constraints per unit of bandwidth.

## III. The Bandwidth Mismatch Case

In this section, we study the problem of bandwidth mismatch between the first and second hop. This problem may arise in many practical situations. For instance, the available bandwidth for the source and the relays to transmit their signals may not be equal. As another example, consider a half-duplex parallel relay channel, assuming a constant bandwidth from the source to the destination, the optimum amount of bandwidth for the first and second hops is not necessarily the same. Hence, the *Combined Amplify-and-Decode Forward (CADF)* scheme is proposed for these types of situations in the sequel.

Here we assume that for each $\rho$ uses of BC channel, one use of the MAC channel is allowed. $\rho$ can be either less or greater than "1". According to the cut-set bound Theorem (See [17]), on the cuts corresponding to the first and second hop, the upper bound, $C_{up}$, on the capacity of this channel, $C_s$, is (See [23]):

$$C_s \leq C_{up} \triangleq \min\left(\rho C\left(MP_s\right), C\left(M^2 P_r\right)\right). \tag{6}$$

### A. The Combined Amplify-and-Decode Forward (CADF)

In this section, CADF scheme is studied. This scheme is illustrated in Figs. 2 and 3. In this strategy, the intended message is split into AF and DF messages. The AF message itself is split into $L$ AF sub-messages. Each AF sub-message is transmitted in $2\alpha_l (l = 1, \cdots, L)$ fraction of the available bandwidth from the source to the destination. The DF message is superimposed on the AF message and transmitted from the source to the relays in $\sum_{l=1}^{L} \alpha_l + \beta_1$ dimensions. Having decoded the DF message, each relay transmits the re-encoded version on top of the AF message in $\sum_{l=1}^{L} \alpha_l + \beta_2$ dimensions (See Fig. 3). Due to the water-filling result of the DF message on the AF message and from (3) and (4), in $\alpha_l$ band from the source to each relay, we have

$$P_{s,AF_l} + P_{s,DF_l} = P_s, \quad l = 1, \cdots, L. \tag{7}$$

Similarly, for the relay side we have

$$P_{r,AF_l} + P_{r,DF_l} = P_r, \quad l = 1, \cdots, L. \tag{8}$$

Furthermore, due to the bandwidth constraint for the BC and MAC channel (See Fig. 3), we have

$$\sum_{l=1}^{L} \alpha_l + \beta_1 = \rho, \tag{9}$$

$$\sum_{l=1}^{L} \alpha_l + \beta_2 = 1. \tag{10}$$

The above discussions result in the following Theorem.

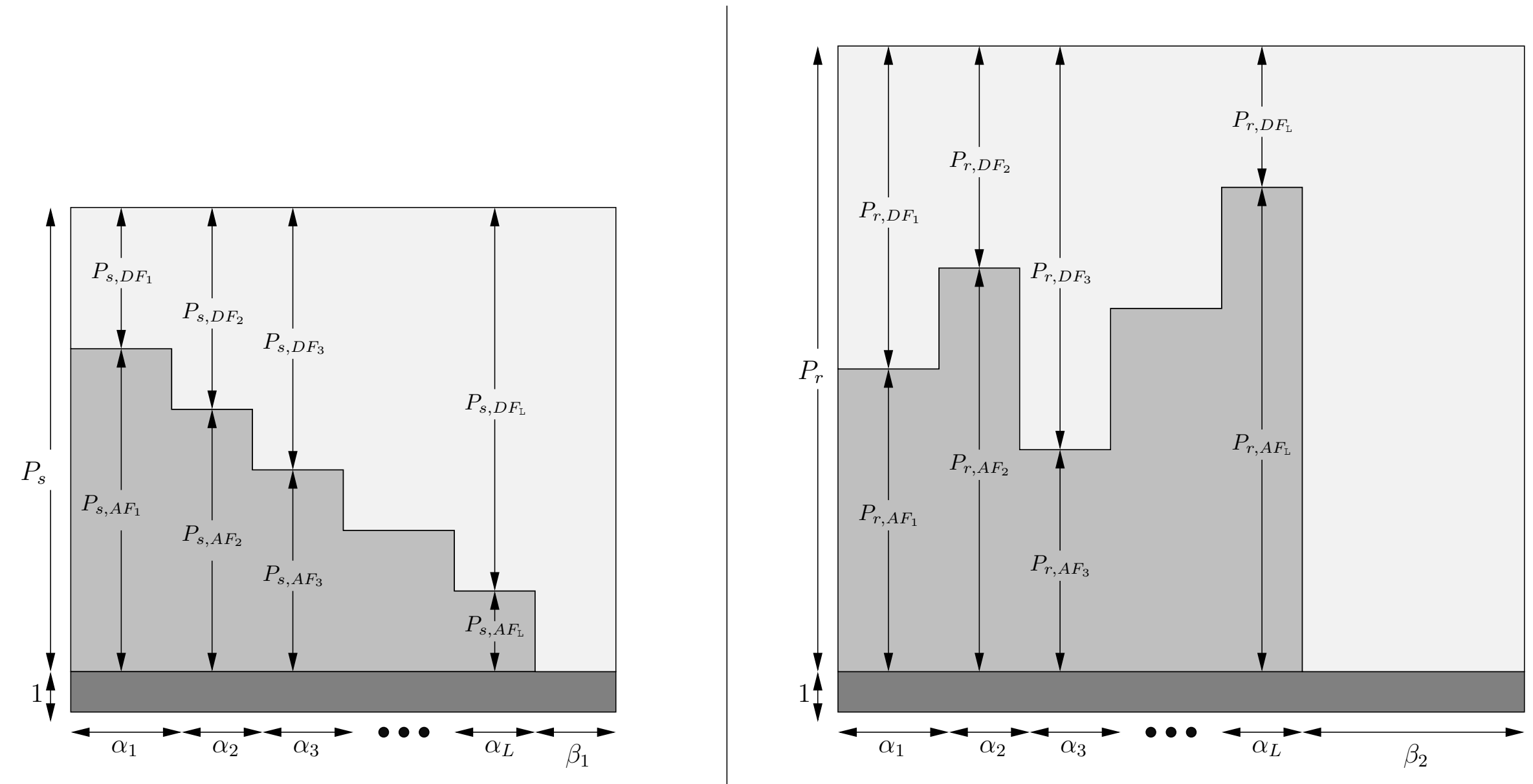

a) Power distribution of the "AF" and "DF" messages at the source side.

b) Power distribution of the "AF" and "DF" messages at the relay side.

Fig. 2. Power distribution of the "AF" and "DF" messages at the source and relay sides.

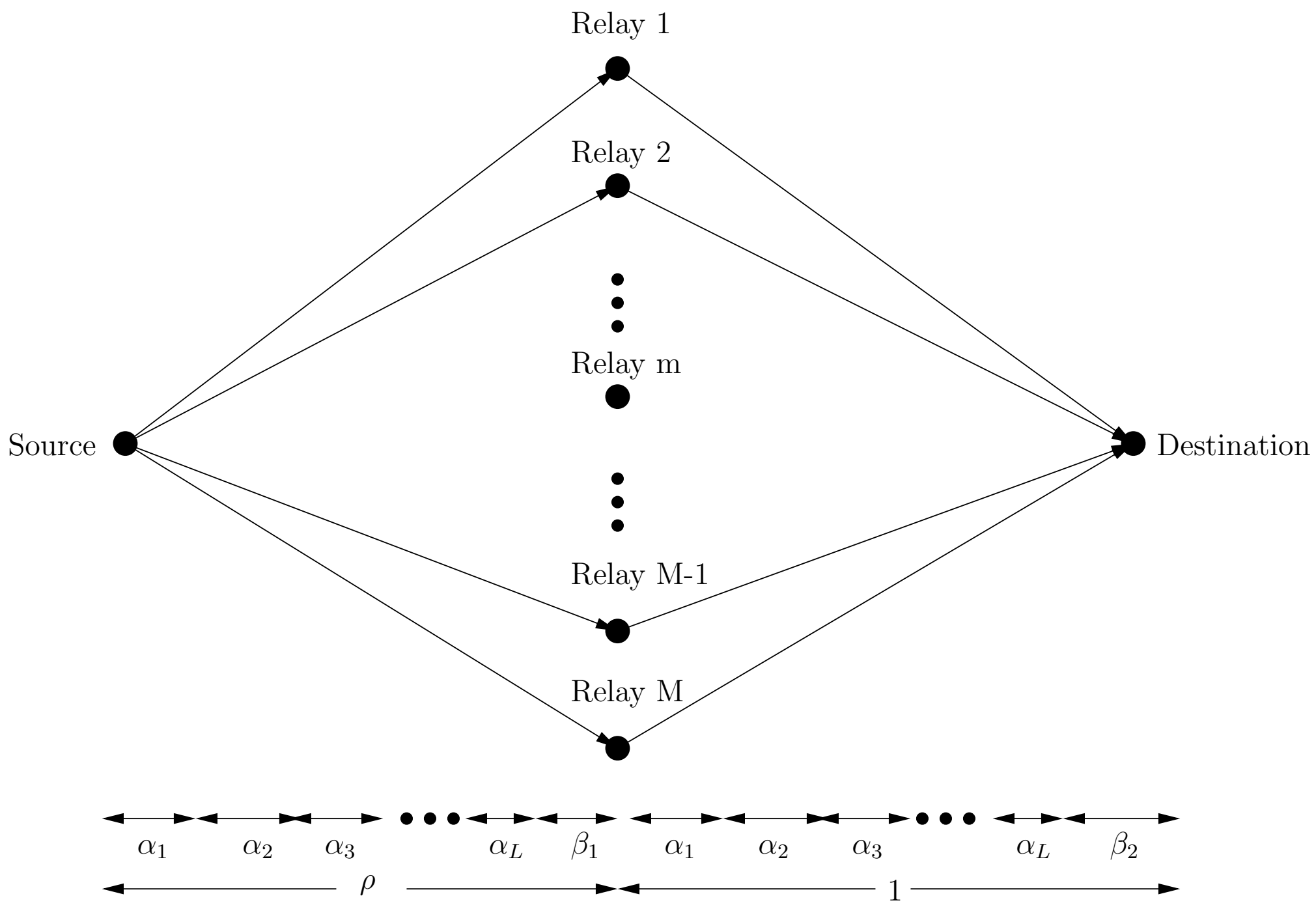

Fig. 3. Bandwidth allocation for the "AF" and "DF" messages for the Gaussian Parallel Relay Channel.

**Theorem 1** *For the Gaussian parallel relay channel, the CADF achieves the following rate:*

$$R_{CADF} \leq \max\min\left(\sum_{l=1}^{L}\alpha_l\left(C\left(\frac{M^2P_{r,AF_l}P_{s,AF_l}}{MP_{r,AF_l}+P_{s,AF_l}+1}\right)+C\left(\frac{P_{s,DF_l}}{P_{s,AF_l}+1}\right)\right)+\beta_1C\left(P_s\right),\right.$$
$$\left.\sum_{l=1}^{L}\alpha_lC\left(\frac{M^2P_rP_{s,AF_l}+M^2P_{r,DF_l}}{MP_{r,AF_l}+P_{s,AF_l}+1}\right)+\beta_2C\left(M^2P_r\right)\right), \tag{11}$$

*subject to:*

$$\sum_{l=1}^{L}\alpha_l+\beta_1=\rho,$$
$$\sum_{l=1}^{L}\alpha_l+\beta_2=1,$$
$$P_{s,AF_l}+P_{s,DF_l}=P_s,$$
$$P_{r,AF_l}+P_{r,DF_l}=P_r,$$
$$0\leq\alpha_l,\beta_1,\beta_2,$$
$$0\leq P_{s,AF_l},P_{s,DF_l}\leq P_s, 0\leq P_{r,AF_l},P_{r,DF_l}\leq P_r, l=1,\cdots,L.$$

*Proof:* See Appendix A. ■

**Remark 1** For the half-duplex scenarios, instead of the constraints $\sum_{l=1}^{L}\alpha_l+\beta_1=\rho$ and $\sum_{l=1}^{L}\alpha_l+\beta_2=1$ for the bandwidths of the first and second hops separately, we assume a constant bandwidth from the source to the destination, i.e., $2\sum_{l=1}^{L}\alpha_l+\beta_1+\beta_2=1$.

**Proposition 1** *The CADF scheme achieves the same rate, assuming successive decoding of the DF and AF messages at the receiver side.*

*Proof:* At band $\alpha_l$ in (11), from Appendix A, we consider the AF and the DF messages as the messages of a MAC with the following inequalities

$$R_{AF_l}\leq\alpha_lC\left(\frac{M^2P_{r,AF_l}P_{s,AF_l}}{MP_{r,AF_l}+P_{s,AF_l}+1}\right), \tag{12}$$
$$R_{DF_l}\leq\alpha_lC\left(\frac{M^2P_{r,DF_l}(P_{s,AF_l}+1)}{MP_{r,AF_l}+P_{s,AF_l}+1}\right), \tag{13}$$
$$R_{AF_l}+R_{DF_l}\leq\alpha_lC\left(\frac{M^2P_rP_{s,AF_l}+M^2P_{r,DF_l}}{MP_{r,AF_l}+P_{s,AF_l}+1}\right). \tag{14}$$

It can be readily verified that subject to the constraint $P_{r,AF_l}+P_{r,DF_l}=P_r$, the right-hand side of (14) is a decreasing function of $P_{r,AF_l}$ or equivalently an increasing function of $P_{r,DF_l}$. Now, let us equate $R_{AF_l}$ in (14) with the AF rate $\acute{R}_{AF_l}$ of another MAC which is achieved by successive decoding of the DF and AF messages. Therefore, we have

$$R_{AF_l} = \acute{R}_{AF_l} = \alpha_l C\left(\frac{M^2 \acute{P}_{r,AF_l} P_{s,AF_l}}{M\acute{P}_{r,AF_l} + P_{s,AF_l} + 1}\right) \leq \alpha_l C\left(\frac{M^2 P_{r,AF_l} P_{s,AF_l}}{M P_{r,AF_l} + P_{s,AF_l} + 1}\right). \quad (15)$$

According to (15), (See Fig. 4) we have

$$\begin{aligned} \acute{P}_{r,AF} &\leq P_{r,AF} \Longrightarrow \\ R_{AF_l} + R_{DF_l} &\leq \acute{R}_{AF_l} + \acute{R}_{DF_l}, \\ R_{DF_l} &\leq \acute{R}_{DF_l}. \end{aligned}$$

Hence, $(R_{AF_l}, R_{DF_l})$ lies in the corner point of the MAC with parameters $(\acute{R}_{AF_l}, \acute{R}_{DF_l})$, i.e. successive decoding of the DF and AF messages achieves $R_{CADF}$. ■

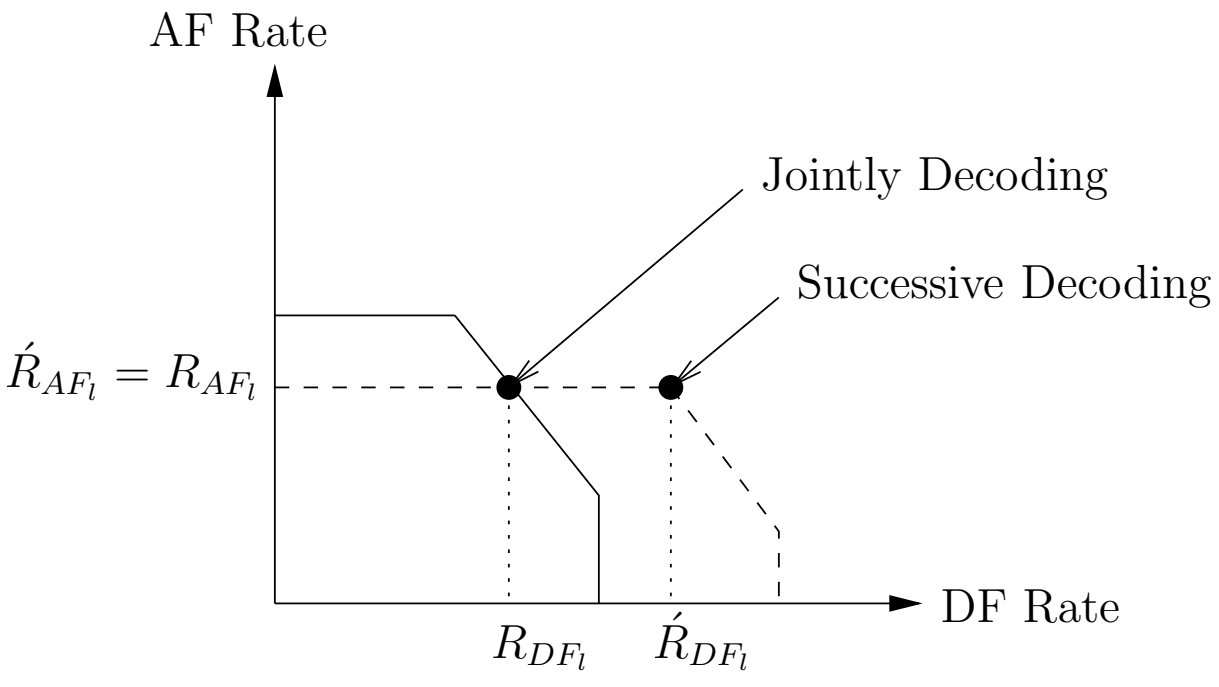

Fig. 4. The order of decoding the DF and AF messages.

**Proposition 2** *The optimum number of bands $L$ in the CADF scheme is at most equal to two. Furthermore, for the half-duplex scenarios assuming one of the $\alpha_l$'s is non-zero, depending on $\rho < 1$ or $\rho > 1$, either $\beta_1 = 0$ and $\beta_2 \neq 0$ or $\beta_1 \neq 0$ and $\beta_2 = 0$.*

*Proof:* Assuming variables $P_{s,AF_l}$, $P_{s,DF_l}$, $P_{r,AF_l}$, and $P_{r,DF_l}$ in (11) as constant parameters, one can cast the optimization problem (11) in a linear form with variables $\alpha_l$, $\beta_1$, and $\beta_2$ as the optimization parameters. In order to do that, we introduce a parameter $\lambda \in \mathbb{R}$ to (11), and assume that the difference between the two terms in the minimization (11) is $\lambda$. Hence, we have the following linear optimization problem which is equivalent to (11):

$$R_{CADF} \leq \max_{\lambda \in \mathbb{R}} \ (\min(-\lambda, 0) + f(\lambda)), \quad (16)$$

where

$$f(\lambda)= \max \sum_{l=1}^{L} \alpha_l \left( C\left(\frac{M^2 P_{r,AF_l} P_{s,AF_l}}{M P_{r,AF_l} + P_{s,AF_l} + 1}\right) + C\left(\frac{P_{s,DF_l}}{P_{s,AF_l}+1}\right)\right) + \beta_1 C\left(P_s\right), \quad (17)$$

subject to:

$$\sum_{l=1}^{L} \alpha_l \left( C\left(\frac{M^2 P_{r,AF_l} P_{s,AF_l}}{M P_{r,AF_l} + P_{s,AF_l} + 1}\right) + C\left(\frac{P_{s,DF_l}}{P_{s,AF_l}+1}\right) - C\left(\frac{M^2 P_r P_{s,AF_l} + M^2 P_{r,DF_l}}{M P_{r,AF_l} + P_{s,AF_l} + 1}\right)\right) + \beta_1 C\left(P_s\right) - \beta_2 C\left(M^2 P_r\right) = \lambda, \quad (18)$$

$$\sum_{l=1}^{L} \alpha_l + \beta_1 = \rho, \quad (19)$$

$$\sum_{l=1}^{L} \alpha_l + \beta_2 = 1, \quad (20)$$

$$0 \leq \alpha_l, \beta_1, \beta_2, l = 1, \cdots, L. \quad (21)$$

For $\rho < 1$, from (19), (20), and knowing $\beta_1 \geq 0$, $\beta_2 > 0$ can be concluded. Hence, substituting $\beta_2$ from (20) into (17) and (18), (17)-(21) becomes

$$f(\lambda)= \max \ \mathbf{c}^T \mathbf{y}, \quad (22)$$

subject to:

$$\mathbf{A}\mathbf{y} = \mathbf{b}, \quad (23)$$

$$\mathbf{y} \succeq 0. \quad (24)$$

where

$$\mathbf{y} = \left[\alpha_1, \alpha_2, \alpha_3, \cdots, \alpha_L, \beta_1\right]^T,$$
$$c_l = C\left(\frac{M^2 P_{r,AF_l} P_{s,AF_l}}{M P_{r,AF_l} + P_{s,AF_l} + 1}\right) + C\left(\frac{P_{s,DF_l}}{P_{s,AF_l}+1}\right), \ l = 1, \cdots, L,$$
$$c_{L+1} = C\left(P_s\right),$$

$$A_{1l} = C\left(\frac{M^2 P_{r,AF_l} P_{s,AF_l}}{M P_{r,AF_l} + P_{s,AF_l} + 1}\right) + C\left(\frac{P_{s,DF_l}}{P_{s,AF_l}+1}\right) - C\left(\frac{M^2 P_r P_{s,AF_l} + M^2 P_{r,DF_l}}{M P_{r,AF_l} + P_{s,AF_l} + 1}\right) + C\left(M^2 P_r\right), \ l = 1, \cdots, L,$$
$$A_{1L+1} = C\left(P_s\right), \ A_{2l} = 1, \ l = 1, \cdots, L+1,$$
$$\mathbf{b} = \left[\lambda + C\left(M^2 P_r\right), \rho\right]^T.$$

The optimum solution of (22), $\mathbf{y}_{opt}$, is an extreme point of the region $\mathcal{F} = \{\mathbf{A}\mathbf{y} = \mathbf{b}, \mathbf{y} \succeq 0\}$. On the other hand, $\mathbf{y}_{opt}$ is an extreme point of $\mathcal{F}$ if and only if it is a basic feasible solution of (22). Since the rank of matrix $\mathbf{A}$ is at most 2, the basic feasible solution of $\mathcal{F}$ has at most 2 non-zero entries (See [25]). Therefore, the only possible cases are $\alpha_i \neq 0$, $\alpha_j \neq 0$ (where $i \neq j$), and $\beta_2 \neq 0$ or $\alpha_i \neq 0$, $\beta_1 \neq 0$, and $\beta_2 \neq 0$.

Having the similar argument for $\rho > 1$, we can easily prove that the only possible cases are $\alpha_i \neq 0$, $\alpha_j \neq 0$ (where $i \neq j$), and $\beta_1 \neq 0$ or $\alpha_i \neq 0$, $\beta_1 \neq 0$, and $\beta_2 \neq 0$. Hence, the optimum number of bands $L$ is at most equal to two.

For the half-duplex scenarios, from Remark 1, the optimization problem (17) becomes a linear optimization problem with two constraints. Using the similar argument as in the bandwidth mismatch case, only two optimization parameters would be non-zero. Hence, assuming one of the $\alpha_l$'s is non-zero and $\rho \neq 1$, depending on $\rho < 1$ or $\rho > 1$, either $\beta_1 = 0$ and $\beta_2 \neq 0$ or $\beta_1 \neq 0$ and $\beta_2 = 0$. Therefore, from the above argument, for the half-duplex scenarios the optimum number of bands $L$ is at most equal to one. ■

By considering the appropriate order of decoding for the DF message and the AF message at the destination and from Proposition 2, the achievable rate can be simplified as

$$R_{CADF} \leq \max \sum_{l=1}^{2} \alpha_l C\left(\frac{M^2 P_{r,AF_l} P_{s,AF_l}}{M P_{r,AF_l} + P_{s,AF_l} + 1}\right) + \min\left(\sum_{l=1}^{2} \alpha_l C\left(\frac{P_{s,DF_l}}{P_{s,AF_l} + 1}\right) + \beta_1 C\left(P_s\right), \right.$$
$$\left. \sum_{l=1}^{2} \alpha_l C\left(\frac{M^2 P_{r,DF_l}(P_{s,AF_l} + 1)}{M^2 P_{r,AF_l} P_{s,AF_l} + M P_{r,AF_l} + P_{s,AF_l} + 1}\right) + \beta_2 C\left(M^2 P_r\right)\right), \tag{25}$$

subject to:

$$\sum_{l=1}^{2} \alpha_l + \beta_1 = \rho, \tag{26}$$

$$\sum_{l=1}^{2} \alpha_l + \beta_2 = 1, \tag{27}$$

$$P_{s,AF_l} + P_{s,DF_l} = P_s, \tag{28}$$

$$P_{r,AF_l} + P_{r,DF_l} = P_r, \tag{29}$$

$$0 \leq \alpha_l, \beta_1, \beta_2, \tag{30}$$

$$0 \leq P_{s,AF_l}, P_{s,DF_l} \leq P_s, 0 \leq P_{r,AF_l}, P_{r,DF_l} \leq P_r,\ l = 1, 2. \tag{31}$$

### *B. The Traditional Coding Schemes*

The achievable rates for the traditional coding schemes such as the Decode-and-Forward (DF), the Amplify-and-Forward (AF), and the Compress-and-Forward (CF) are derived in [23]. These are highlighted for comparison purposes:

*1) Decode-and-Forward (DF):* In this scheme, the codeword $\mathbf{x}_m$ in (2) is a re-encoded version of the decoded message at relay $m$. Hence, the source transmits its message such that each relay can decode it. Hence, the DF scheme achieves

$$R_{DF} = \min\left(\rho C\left(P_s\right), C\left(M^2 P_r\right)\right). \tag{32}$$

*2) Amplify-and-Forward (AF):* In the AF scheme, the relay $m$ transmits a re-scaled version of the signal received from the BC channel. Hence, the AF scheme achieves

$$R_{AF} = \gamma C\left(\frac{M^2 P_r P_s}{M P_r + P_s + 1}\right). \tag{33}$$

where $\gamma = \min(\rho, 1)$.

*3) Compress-and-Forward (CF):* In the CF scheme, the relay $m$ estimates the transmitted codeword and digitally compresses its estimation. Then, it encodes the compressed value to an appropriate channel codeword and sends it over the MAC channel [23]. Hence, the CF scheme achieves

$$R_{CF}= \rho C\left(P_{CF}\right), \tag{34}$$

subject to:

$$(1+MP_r)^{\frac{1}{\rho}} = 1 + P_{CF}\left(\frac{MP_s}{MP_s - P_{CF}+1}\right)^M .$$

### C. The Rematch-and-Forward (RF) scheme

The RF scheme can be briefly explained as follows. Depending on $\rho > 1$ or $\rho < 1$, the source conducts the up-sampling or down-sampling operation, and the relays do the reverse operation and then estimate the transmitted signal. Indeed, this scheme matches a colored source to a channel and is implemented using the modulo lattice operation. For further details see [21] [22] [23]. The following Theorem is proved in [23].

**Theorem 2** *For the Gaussian parallel relay channel with expansion factor $\rho$, assuming $P_s > 1$, the RF scheme achieves the following rate*

$$R_{RF} = C\left(\frac{M^2P_r(P_s^\rho - 1)}{(P_s^\rho + MP_r)^\gamma (P_s^\rho + M^2P_r)^{1-\gamma}}\right). \tag{35}$$

**Theorem 3** *The CADF scheme achieves a better rate than the RF scheme, i.e., $R_{CADF} \geq R_{RF}$.*

*Proof:* Throughout the proof, we assume that $L = 1$ and depending on $\rho < 1$ or $\rho > 1$, either $\beta_1 = 0$ and $\beta_2 \neq 0$ or $\beta_1 \neq 0$ and $\beta_2 = 0$.

*Case 1 :* $\rho \leq 1$

Consider the proposed scheme with $P_{s,AF} = P_s^\rho - 1$, $P_{s,DF} = P_s - P_s^\rho + 1$, and assume that no DF message is superimposed on the AF message at the relay, i.e. $P_{r,AF} = P_r$ and $P_{r,DF} = 0$. Hence, the achievable rate of the CADF scheme can be simplified to

$$R_{CADF} = \rho C\left(\frac{M^2P_r\left(P_s^\rho - 1\right)}{MP_r + P_s^\rho}\right) + \min\left\{\rho C\left(\frac{P_s - P_s^\rho + 1}{P_s^\rho}\right), (1-\rho)C(M^2P_r)\right\} \tag{36}$$

Now, let us define $SNR_{AF} \triangleq \frac{M^2P_r\left(P_s^\rho - 1\right)}{MP_r + P_s^\rho}$ and $SNR_{KF} \triangleq \frac{M^2P_r\left(P_s^\rho - 1\right)}{P_s^\rho + M^2P_r}$. It is easy to show that

$$R_{CADF} \geq \rho C(SNR_{AF}) + (1-\rho)C(SNR_{KF}). \tag{37}$$

To prove this, consider the fact that $SNR_{KF} \leq M^2 P_r$ and on the other hand, since $P_s > 1$ as in [23], we have $\left(\frac{P_s+1}{P_s}\right)^{\rho}\left(\frac{P_s^{\rho}+M^2P_r}{1+M^2P_r}\right)^{1-\rho} \geq 1$ which results in $(1-\rho)\log\left(\frac{P_s^{\rho}\left(1+M^2P_r\right)}{P_s^{\rho}+M^2P_r}\right) \leq \rho\log\left(\frac{P_s+1}{P_s^{\rho}}\right)$ or equivalently $(1-\rho)C(SNR_{KF}) \leq \rho C\left(\frac{P_s-P_s^{\rho}+1}{P_s^{\rho}}\right)$. Now, we can lower-bound the right-hand-side of (37) as follows

$$\begin{aligned}
\rho C(SNR_{AF}) + (1-\rho)C(SNR_{KF}) &= \rho\log(1+SNR_{AF}) + (1-\rho)\log(1+SNR_{KF}) \\
&= \log\left((1+SNR_{AF})^{\rho}(1+SNR_{KF})^{1-\rho}\right) \\
&\overset{(a)}{\geq} \log\left(1+SNR_{AF}^{\rho}SNR_{KF}^{1-\rho}\right) \\
&= R_{RF}. \qquad (38)
\end{aligned}$$

Here, $(a)$ follows from applying Holder's inequality with $p = \frac{1}{\rho}$ and $q = \frac{1}{1-\rho}$ (See [24]). Comparing (37) and (38) completes the proof.

*Case 2 :* $\rho \geq 1$

For the sake of simplicity we assume that no DF message is superimposed on the AF message at the source, i.e. $P_{s,AF} = P_s$ and $P_{s,DF} = 0$. Here two cases are considered:

i) $(\rho-1)C(P_s) > C(M^2P_r)$. In this case, we have $R_{CADF} = R_{DF} = C(M^2P_r)$ which is obviously greater than $R_{RF}$. In fact, $R_{CADF}$ is also equal to the capacity of the channel.

ii) otherwise, we have

$$R_{CADF} = C\left(\frac{M^2\left(P_{r,AF}+P_{r,DF}\right)P_s}{MP_{r,AF}+P_s}\right), \qquad (39)$$

where re-scaling the AF portion of the received signal at the relay with $\sqrt{\frac{P_{r,AF}}{P_s}}$, we have $P_{r,AF} + P_{r,DF} + \frac{P_{r,AF}}{P_s} = P_r$. Simplifying (39), we have

$$R_{CADF} = C\left(\frac{MP_s\left(1+MP_r\right)}{MP_{r,AF}+P_s} - M\right), \qquad (40)$$

On the other hand, knowing

$$(\rho-1)C(P_s) = C\left(\frac{M^2P_{r,DF}}{M^2P_{r,AF}+\frac{MP_{r,AF}}{P_s}+1}\right), \qquad (41)$$

we can derive $P_{r,AF}$ as

$$MP_{r,AF} = \frac{M^2P_sP_r - P_s^{\rho}}{MP_s^{\rho}+P_s^{\rho-1}+MP_s+M}. \qquad (42)$$

From (42), one can easily verify that $MP_{r,AF} < \frac{MP_r}{P_s^{\rho-1}}$. Substituting $MP_{r,AF}$ with $\frac{MP_r}{P_s^{\rho-1}}$ in (40), we conclude that $R_{CADF} > R_{RF}$. ■

## IV. Simulation Results

In this section, the achievable rates of the proposed CADF scheme with that of the traditional coding schemes and the upper bound are compared.

Fig. 5 compares the achievable rates of different schemes when $\rho = 0.5 < 1$. On the other hand, Fig. 6 compares the achievable rates of different schemes when $\rho = 2 > 1$. As we proved in the previous sections and, from these figures, as the number of relays increases, the CADF scheme always outperforms the RF scheme.

Figs. 7 and 8 compare the achievable rate of the CADF scheme with that of other schemes for the half-duplex scenarios. Assuming a constant bandwidth from the source to the destination, the optimum bandwidths for the first and second hops are obtained. Fig. 7 show that, as the number of relays increases, the CADF scheme outperforms the other schemes considerably. On the other hand, from Fig. 8, although the CADF scheme gives a better achievable rate compared to the RF scheme, it eventually coincides with the AF scheme.

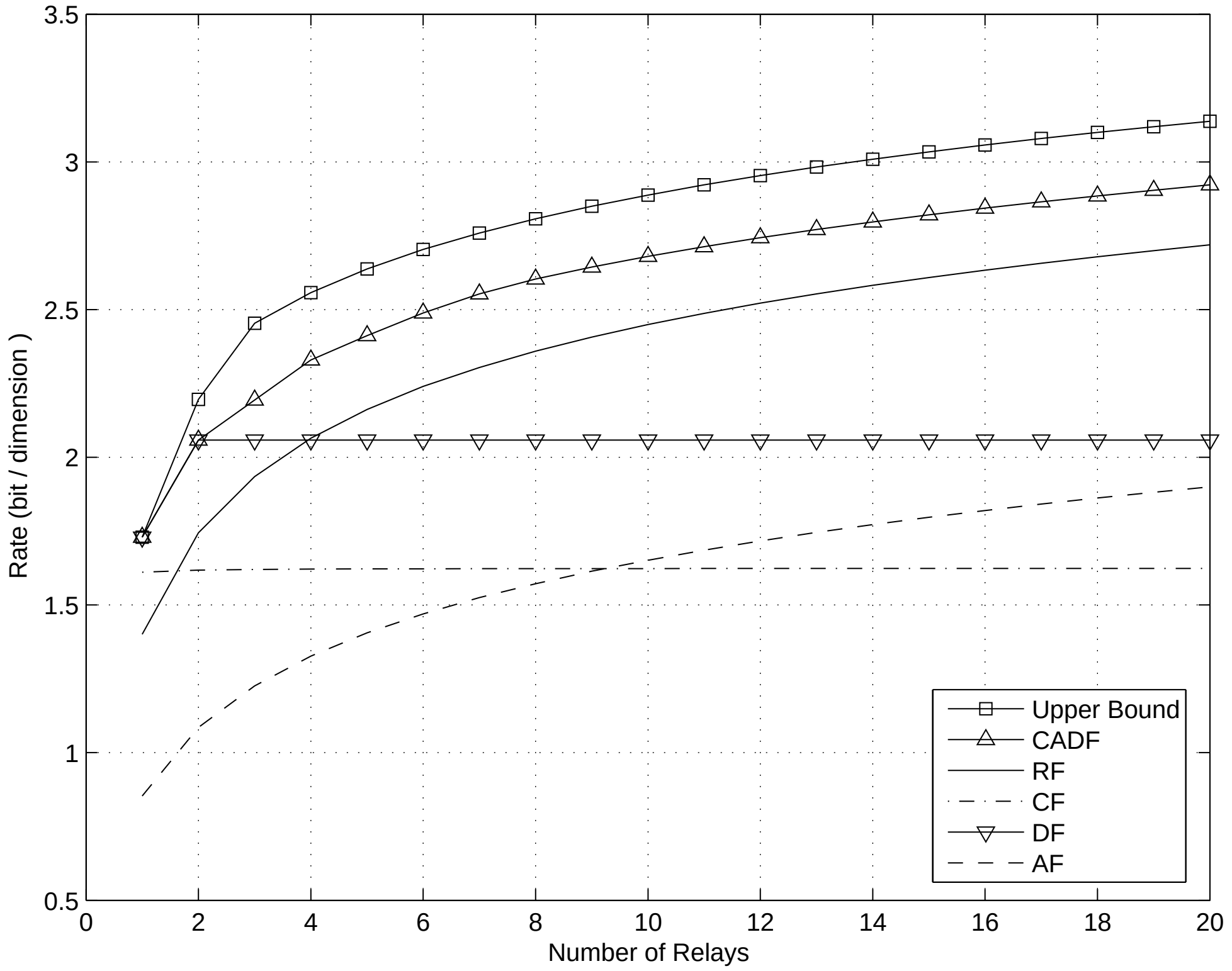

Fig. 5. Rate versus number of relays ($\rho = 0.5$, $P_s = 300$, $MP_r = 10$).

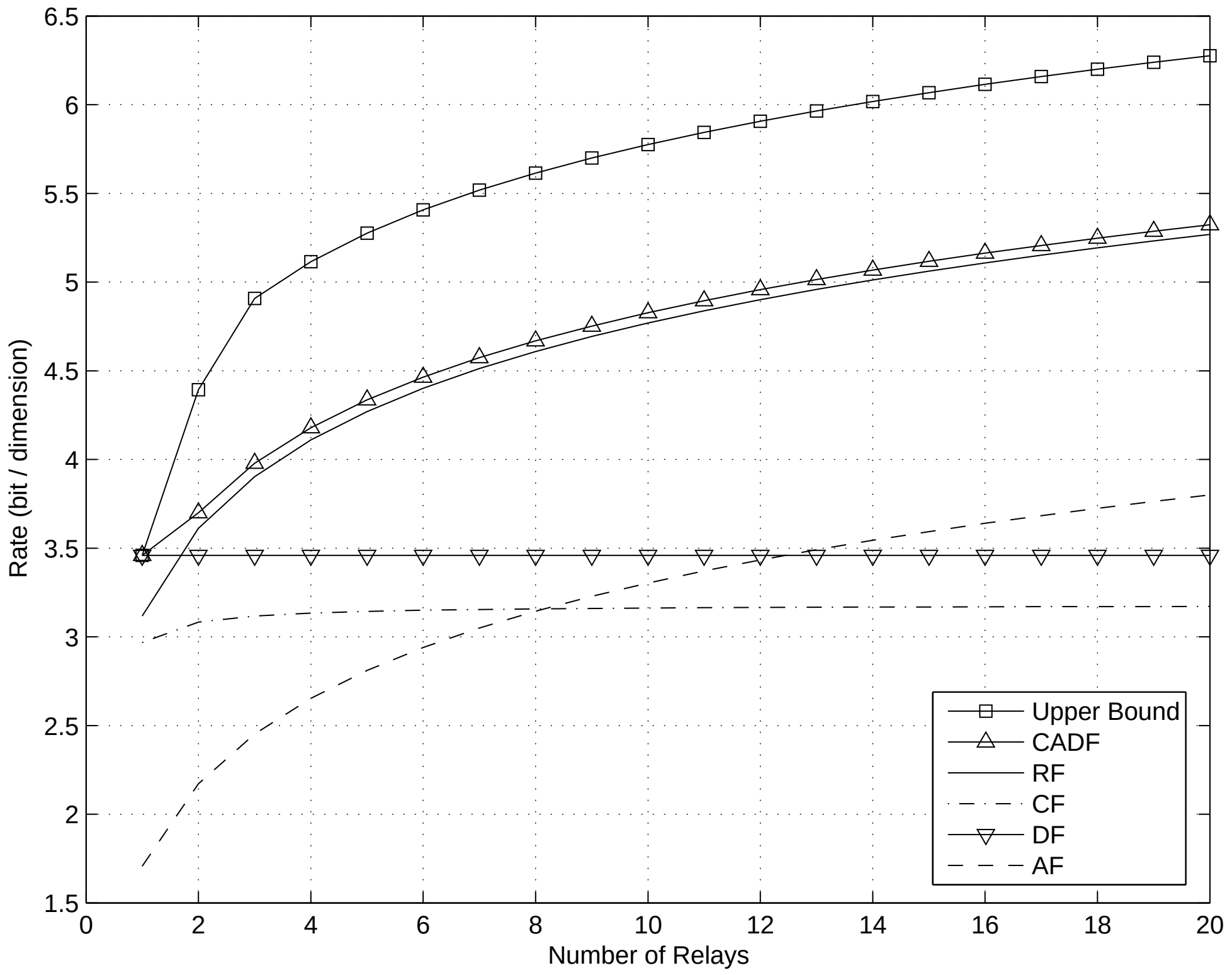

Fig. 6. Rate versus number of relays ($\rho = 2$, $P_s = 10$, $MP_r = 300$).

Fig. 9 compares the achievable rate of the CADF-DF with that of the RF-DF in [23], and the AF-DF of [8] [9] in Schein's parallel relay setup (i.e. parallel relay with two relays and no bandwidth mismatch). Here, we assume that $P_s = 20(dB)$. In this figure, we assume that the total dimensions from the source to the destination is "2". The assigned dimension to the BC channel is equal to the one assigned to the MAC channel. In the time sharing between the CADF and DF schemes, $t_1 + t_2$ dimensions are assigned to the CADF scheme ($t_1$ dimensions for the BC channel, and $t_2$ dimensions for the MAC channel) while $2 - t_1 - t_2$ is assigned to the DF scheme ($1 - t_1$ dimensions for the BC channel, and $1 - t_2$ dimensions for the MAC channel) with different peak powers. The same time sharing pattern is used for the time sharing between the RF and the DF schemes [23].

As Fig. 9 shows, the CADF-DF considerably outperforms the RF-DF and AF-DF. It is worth noting that as the Schein's AF-DF can be considered as a special case of the CADF-DF, we can expect that the achievable rate of the CADF-DF is always better than the AF-DF. On the other hand, from the result of Theorem 3, the CADF-DF always outperforms the RF-DF in the

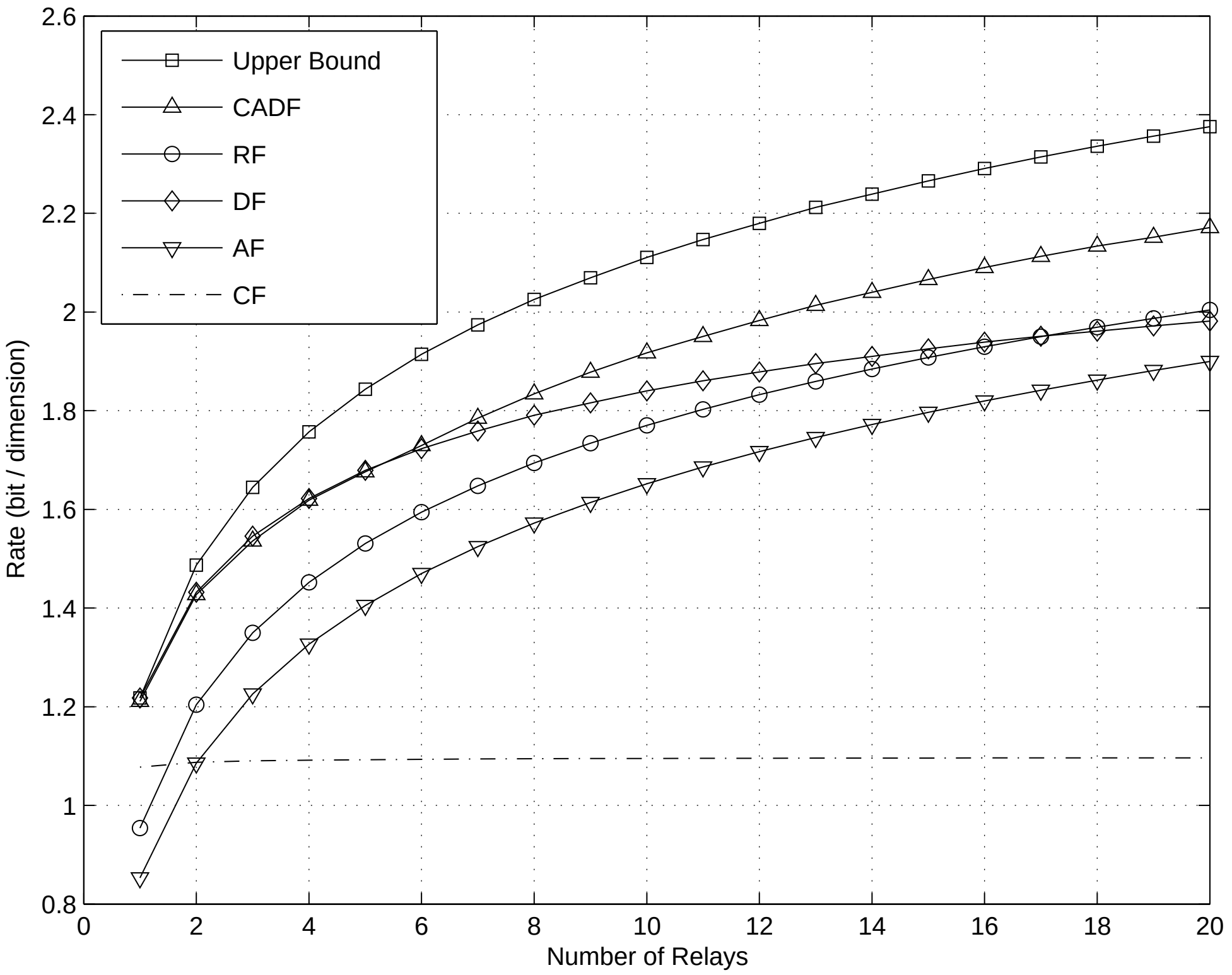

Fig. 7. Rate versus number of relays for the half-duplex scenario ($P_s = 300$, $MP_r = 10$).

Schein's setup.

## V. Conclusion

This paper considered the problem of data transmission for the Gaussian parallel relay channel when there is a bandwidth mismatch between the BC channel and the MAC channel. A *Combined Amplify-and-Decode Forward* (CADF) scheme was proposed and it was proved that the CADF always outperforms the RF scheme presented in [23]. It was also shown that the CADF scheme always outperforms other traditional coding schemes, i.e., AF, DF, and CF. For the case in which there exists no bandwidth mismatch between the BC and the MAC channels, using the time sharing between the CADF and DF schemes (CADF-DF) always outperforms the RF-DF in [23], and the AF-DF in [8] [9].

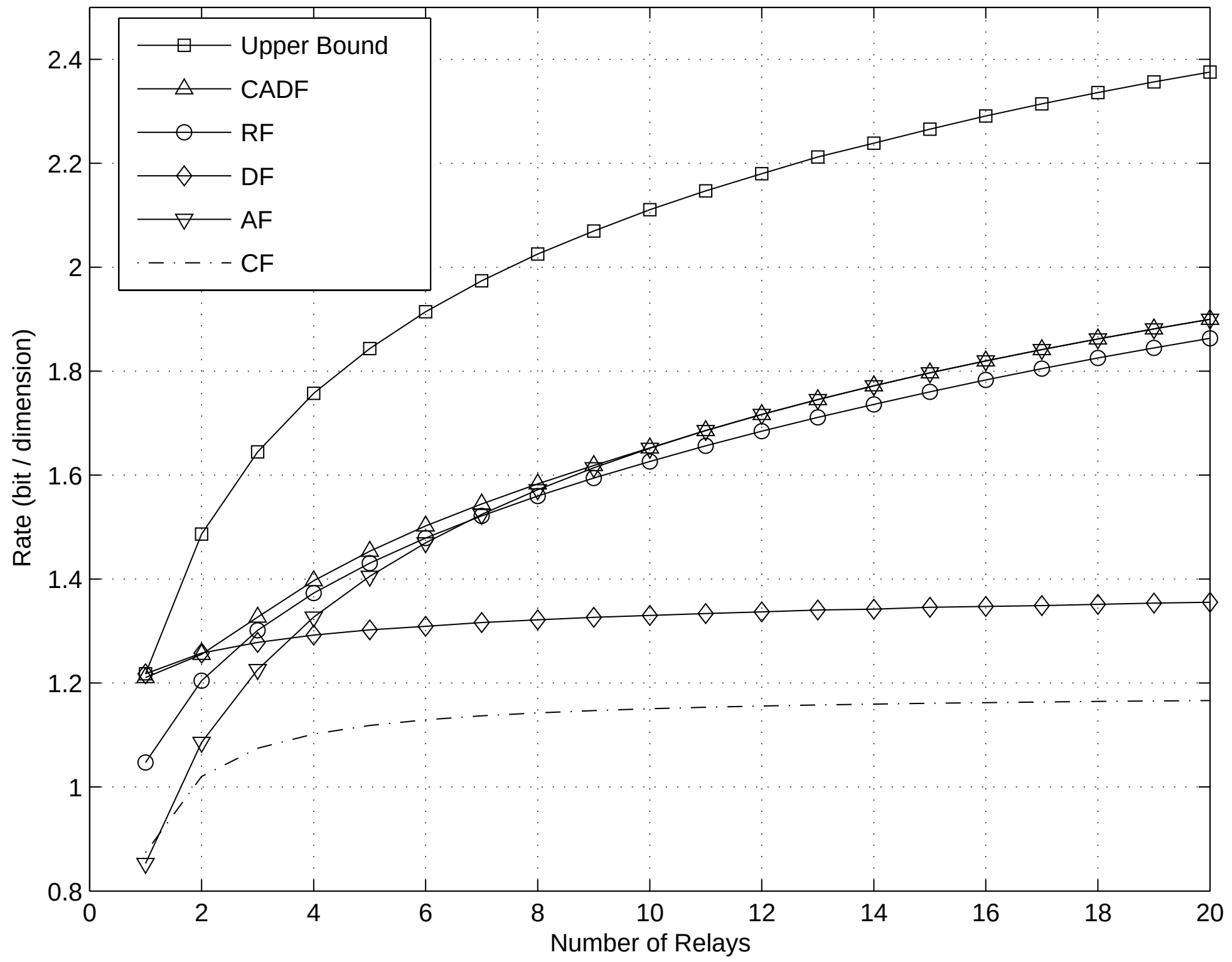

Fig. 8. Rate versus number of relays for the half-duplex scenario ($P_s = 10$, $MP_r = 300$).

# Appendix A

## *Proof of Theorem 1*

*Codebook Construction:*

At band $\alpha_l$, $(l = 1, \cdots, L)$ and $\beta_1$, the source generates $2^{nR_{AF_l}}$, $2^{nR_{DF_l}}$, and $2^{nR_{DF}}$ sequences $\mathbf{v}_{BC_l}(w_{AF_l})$, $\mathbf{u}_{BC_l}(w_{DF_l})$, and $\mathbf{x}_{BC}(w_{DF})$ according to $\prod_{i=1}^{\alpha_l n} p(v_{BC_l,i})$, $\prod_{i=1}^{\alpha_l n} p(u_{BC_l,i})$, and $\prod_{i=1}^{\beta_1 n} p(x_{BC,i})$, respectively. $V_{BC_l}$, $U_{BC_l}$, and $X_{BC}$ are Gaussian random variables with zero mean and variances $P_{s,AF_l}$, $P_{s,DF_l}$, and $P_s$ per dimension, where $P_{s,AF_l} + P_{s,DF_l} = P_s$. Furthermore, at band $\alpha_l$, the source generates i.i.d sequences $\mathbf{x}_{BC_l}$, where we have $X_{BC_l} = V_{BC_l} + U_{BC_l}$. Hence, $X_{BC_l} \sim \mathcal{N}(0, P_s)$.

All the relays, at band $\alpha_l$, $(l = 1, \cdots, L)$, and $\beta_2$ generate $2^{nR_{DF_l}}$ and $2^{nR_{DF}}$ i.i.d $\mathbf{u}_{r_l}(w_{DF_l})$, and $\mathbf{x}_r(w_{DF})$ sequences according to probabilities $\prod_{i=1}^{\alpha_l n} p(u_{r_l,i})$, and $\prod_{i=1}^{\beta_2 n} p(x_{r,i})$. $U_{r_l}$ and $X_r$ are Gaussian random variables with zero mean and variances $P_{r,DF_l}$ and $P_r$ per dimension.

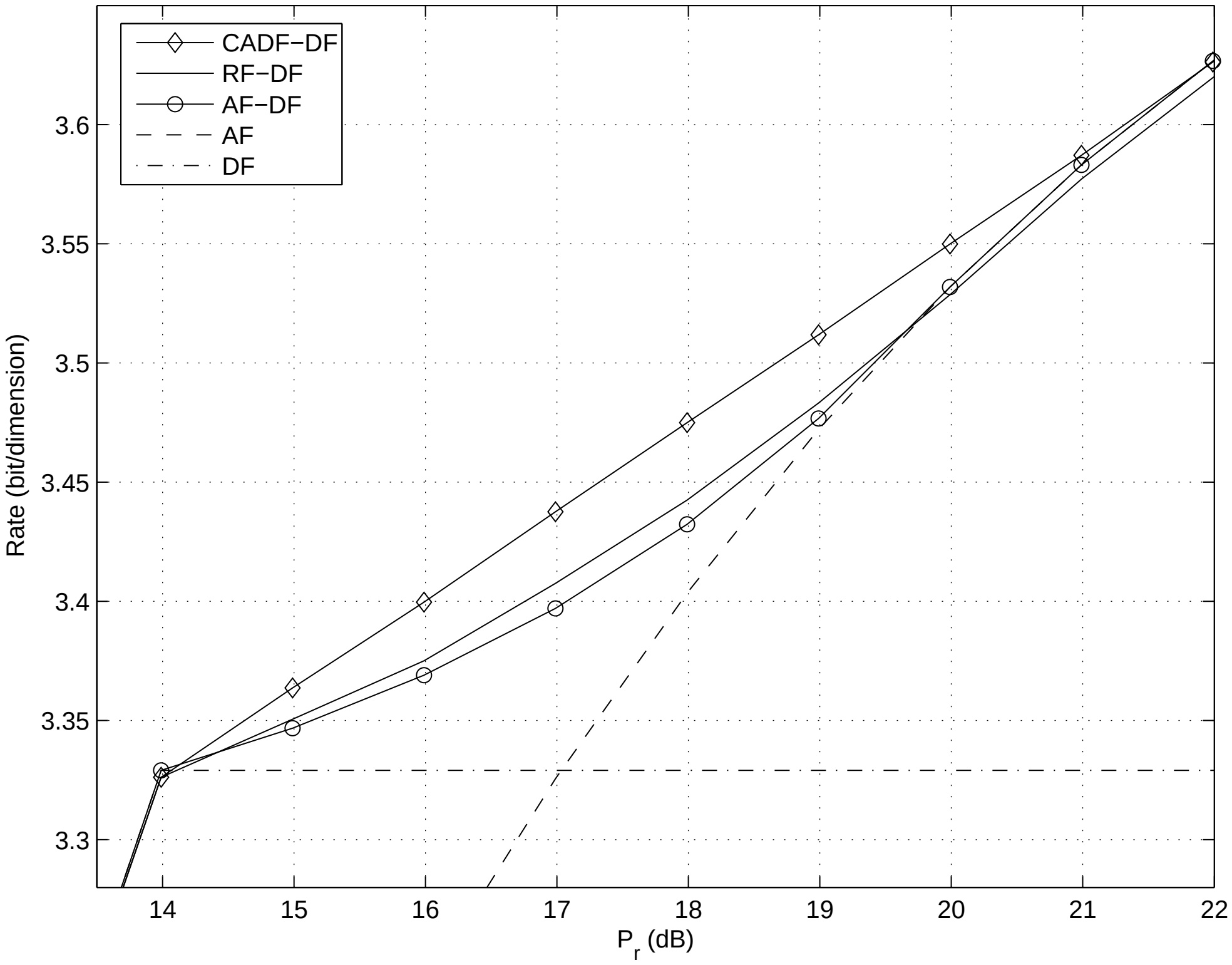

Fig. 9. Achievable Rates by Time Sharing.

Furthermore, relay $m$ generates i.i.d sequences $\mathbf{x}_{m_l}$, due to

$$X_{m_l} = \sqrt{\frac{P_{r,AF_l}}{P_{s,AF_l}+1}}(V_{BC_l}+Z_m)+U_{r_l}. \tag{43}$$

*Encoding:*

*Encoding at the source:*

At band $\alpha_l$, the source encodes $w_{AF_l} \in \{1,\cdots,2^{nR_{AF_l}}\}$, and $w_{DF_l} \in \{1,\cdots,2^{nR_{DF_l}}\}$ to $\mathbf{v}_{BC_l}(w_{AF_l})$ and $\mathbf{u}_{BC_l}(w_{DF_l})$ and sends $\mathbf{x}_{BC_l}(w_{AF_l}, w_{DF_l})$ to the relays. Furthermore, at band $\beta_1$, the source encodes $w_{DF} \in \{1,\cdots,2^{nR_{DF}}\}$ to $\mathbf{x}_{BC}(w_{DF})$ and sends it to the relays.

*Encoding at relay $m$:*

At band $\alpha_l$, relay $m$ encodes $w_{DF_l} \in \{1,\cdots,2^{nR_{DF_l}}\}$ to $\mathbf{u}_{r_l}(w_{DF_l})$ and sends $\mathbf{x}_{m_l}$ as obtained in (43), to the destination. Furthermore, at band $\beta_2$, relay $m$ encodes $w_{DF} \in \{1,\cdots,2^{nR_{DF}}\}$ to $\mathbf{x}_r(w_{DF})$ and sends it to the destination.

*Decoding:*

*Decoding at relay $m$:*

At band $\alpha_l$, relay $m$ declares $\hat{w}_{DF_l} = w_{DF_l}$ iff there exits a unique $\mathbf{u}_{BC_l}(w_{DF_l})$, such that $\left(\mathbf{u}_{BC_l}(w_{DF_l}), \mathbf{y}_{m_l}\right) \in A_\epsilon^{(n)}$ (See [17]). Hence, in order to make the probability of error zero, we have

$$R_{DF_l} \leq \alpha_l C\left(\frac{P_{s,DF_l}}{P_{s,AF_l}+1}\right). \tag{44}$$

Similarly, at band $\beta_1$, relay $m$ declares $\hat{w}_{DF} = w_{DF}$ iff there exits a unique $\mathbf{x}_{BC}(w_{DF})$, such that $(\mathbf{x}_{BC}(w_{DF}), \mathbf{y}_m) \in A_\epsilon^{(n)}$. Hence, in order to make the probability of error zero, we have

$$R_{DF} \leq \beta_1 C(P_s). \tag{45}$$

*Decoding at the final destination:*

At band $\alpha_l$, the destination declares $\hat{w}_{AF_l} = w_{AF_l}$ and $\hat{w}_{DF_l} = w_{DF_l}$ iff there exits unique $\mathbf{v}_{BC_l}(w_{AF_l})$ and $\mathbf{u}_{r_l}(w_{DF_l})$, such that $\left(\mathbf{v}_{BC_l}(w_{AF_l}), \mathbf{u}_{r_l}(w_{DF_l}), \mathbf{y}_{MAC_l}\right) \in A_\epsilon^{(n)}$. Hence, in order to make the probability of error zero, we have

$$R_{AF_l} \leq \alpha_l C\left(\frac{M^2 P_{r,AF_l} P_{s,AF_l}}{M P_{r,AF_l} + P_{s,AF_l} + 1}\right), \tag{46}$$

$$R_{DF_l} \leq \alpha_l C\left(\frac{M^2 P_{r,DF_l}(P_{s,AF_l}+1)}{M P_{r,AF_l} + P_{s,AF_l} + 1}\right), \tag{47}$$

$$R_{AF_l} + R_{DF_l} \leq \alpha_l C\left(\frac{M^2 P_r P_{s,AF_l} + M^2 P_{r,DF_l}}{M P_{r,AF_l} + P_{s,AF_l} + 1}\right). \tag{48}$$

However, as indicated in Proposition 1 the same rate $R_{CADF}$ is achievable by successive decoding of the DF and AF messages, hence, we can assume

$$R_{AF_l} = \alpha_l C\left(\frac{M^2 P_{r,AF_l} P_{s,AF_l}}{M P_{r,AF_l} + P_{s,AF_l} + 1}\right). \tag{49}$$

Now, from (48) and (49) inequality (47) is concluded. Hence, inequality (47) is extra.

Similarly at band $\beta_2$, destination declares $\hat{w}_{DF} = w_{DF}$ iff there exits a unique $\mathbf{x}_r(w_{DF})$, such that $(\mathbf{x}_r(w_{DF}), \mathbf{y}_{MAC}) \in A_\epsilon^{(n)}$. Hence, in order to make the probability of error zero, we have

$$R_{DF} \leq \beta_2 C\left(M^2 P_r\right). \tag{50}$$

Noting the fact that $R_{CADF} = \sum_{l=1}^{L}(R_{AF_l} + R_{DF_l}) + R_{DF}$, and from (44), (45), (48), (49), and (50), Theorem 1 is proved.